\documentclass[acmsmall,screen]{acmart}

\AtBeginDocument{%
  }
\setcopyright{cc}
\setcctype{by-nc-nd}
\acmJournal{PACMCGIT}
\acmYear{2026} 
\acmVolume{9} 
\acmNumber{3} 
\acmArticle{47}
\acmMonth{7} 
\acmDOI{10.1145/3816091}

\acmSubmissionID{121}

\begin{document}

\title{Printing the Underdetermined: Materializing Multi-solutionness in Figurative Paintings}

\author{Yutao Ming}
\authornote{Both authors contributed equally to this research.}
\email{mingyt2025@shanghaitech.edu.cn}
\orcid{0009-0001-3220-7802}
\affiliation{%
  \institution{ShanghaiTech University}
  \city{Shanghai}
  \country{China}
}
\affiliation{%
  \institution{Crysta AI}
  \city{Shanghai}
  \country{China}
}

\author{Teng Xu}
\authornotemark[1]
\email{xt@shanghaitech.edu.cn}
\orcid{0000-0003-4747-2219}
\affiliation{%
  \institution{ShanghaiTech University}
  \city{Shanghai}
  \country{China}
}
\affiliation{%
  \institution{Crysta AI}
  \city{Shanghai}
  \country{China}
}

\author{Youjia Wang}
\email{wangyj2@shanghaitech.edu.cn}
\orcid{0000-0002-0517-3475}
\affiliation{%
  \institution{ShanghaiTech University}
  \city{Shanghai}
  \country{China}
}
\affiliation{%
  \institution{Crysta AI}
  \city{Shanghai}
  \country{China}
}

\author{Yunyang Liu}
\email{liuyy2024@shanghaitech.edu.cn}
\orcid{0009-0001-9250-2512}
\affiliation{%
  \institution{ShanghaiTech University}
  \city{Shanghai}
  \country{China}
}
\affiliation{%
  \institution{Crysta AI}
  \city{Shanghai}
  \country{China}
}

\author{Fengmin Yang}
\email{yangfm2023@shanghaitech.edu.cn}
\orcid{0009-0009-9864-2944}
\affiliation{%
  \institution{ShanghaiTech University}
  \city{Shanghai}
  \country{China}
}
\affiliation{%
  \institution{Crysta AI}
  \city{Shanghai}
  \country{China}
}

\author{Fuqiang Zhao}
\email{zhaofq@crysta.ai}
\orcid{0000-0003-2786-5699}
\affiliation{%
  \institution{Crysta AI}
  \city{Shanghai}
  \country{China}
}

\author{Jingyi Yu}
\email{yujingyi@shanghaitech.edu.cn}
\orcid{0000-0002-8580-0036}
\affiliation{%
  \institution{ShanghaiTech University}
  \city{Shanghai}
  \country{China}
}

\author{Hua Yang}
\email{yujingyi@shanghaitech.edu.cn}
\orcid{0009-0004-5211-4744}
\affiliation{%
  \institution{ShanghaiTech University}
  \city{Shanghai}
  \country{China}
}

\author{Yanjun Zhou}
\email{zhouyj5@shanghaitech.edu.cn}
\orcid{0009-0005-0838-738X}
\affiliation{%
  \institution{ShanghaiTech University}
  \city{Shanghai}
  \country{China}
}

\renewcommand{\shortauthors}{Ming, Xu, et al.}

\begin{abstract}

Figurative paintings are often approached as if they depict a single recoverable 3D scene: viewers infer depth and occlusion, and reconstruction pipelines attempt to converge to one stable model.
We instead foreground \emph{multi-solutionness}, the non-uniqueness of 3D configurations compatible with a single painted image, and propose a workflow that keeps this non-uniqueness visible and material.
Multi-solutionness arises from two sources: \emph{unobserved content}, where backsides and occluded volumes admit multiple plausible completions, and \emph{observed cues}, where perspective, shading, and occlusion still underconstrain geometry.
When additional views are synthesized by a video generative model without explicit 3D constraints, small frame-level drifts become inevitable rather than exceptional.
Our pipeline samples multiple camera-orbit multi-view video sequences from one painting, reconstructs each sequence with 3D Gaussian Splatting into a point-based Gaussian scene representation where density halos and ghosting expose unresolved degrees of freedom, and fabricates these representations as physical artifacts using DreamPrinting.
By treating multiple compatible interpretations as explicit outputs rather than residual error, we provide a computational framework for spatial readings of figurative painting that can be inspected, compared, and discussed in both digital and physical form.

\end{abstract}




\begin{teaserfigure}
  \centering
  \includegraphics[width=0.8\linewidth]{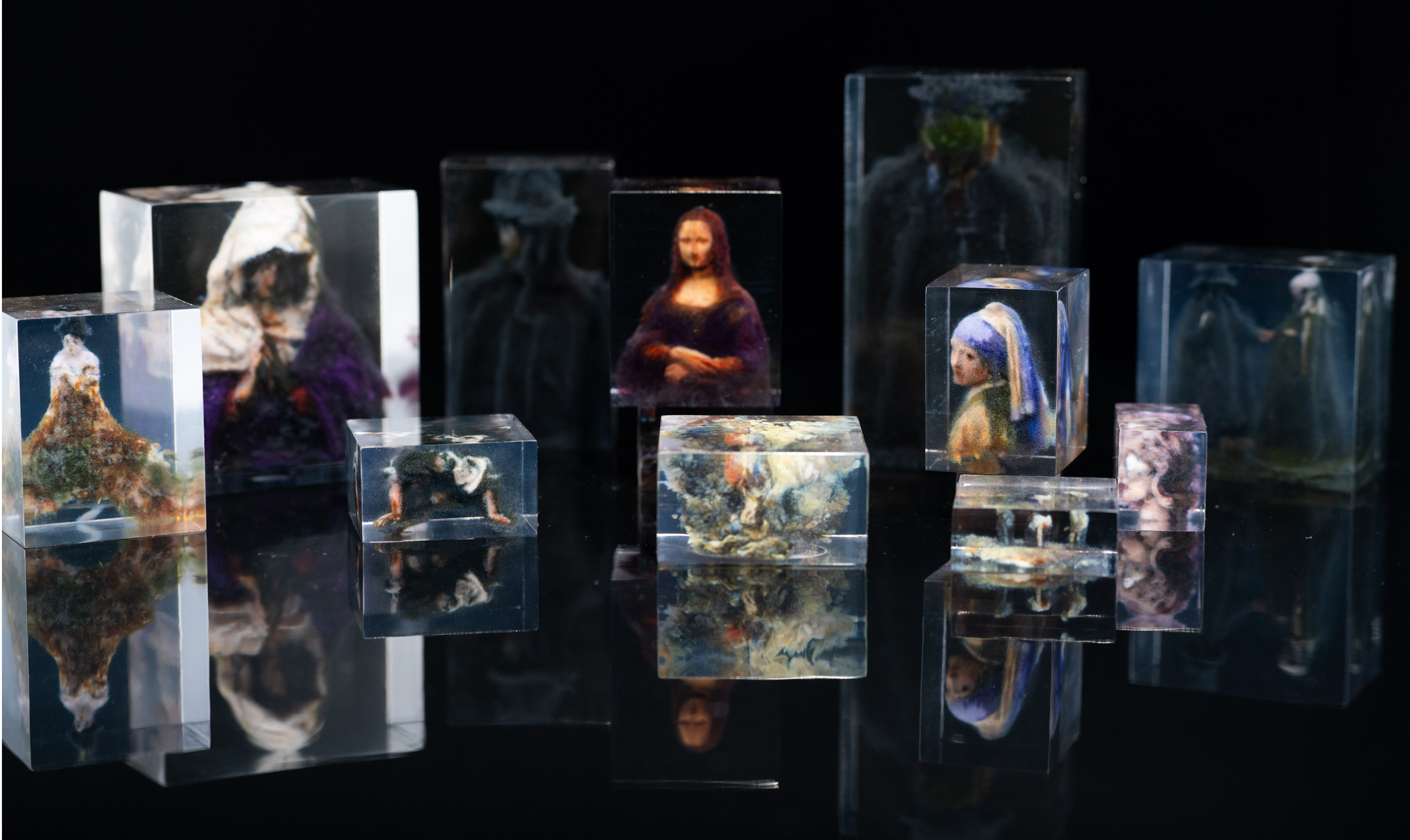}
\caption{\textbf{Teaser.} A set of volumetric prints materializing multi-solutionness from figurative paintings. Each object is fabricated from a hypothesis-conditioned 3D Gaussian Splatting reconstruction generated from a single painting, preserving alternative unobserved-content completions across objects and boundary diffusion/ghosting within each object as traces of underconstrained observed cues.}
  \label{fig:teaser}
\end{teaserfigure}

\maketitle

\section{Introduction}
Figurative paintings are routinely approached as if they were windows onto coherent worlds.
In everyday viewing, pictorial cues, including perspective, occlusion, and shading, invite viewers to infer depth and spatial relations beyond the canvas.
In computer vision and graphics, the same cues are treated as evidence from which a 3D representation can be recovered.
Across both perception and computation, the implied goal is often a single, stable scene: a definitive ``what is really there.''

Yet pictorial space is structurally underdetermined.
A single painted image can remain compatible with multiple 3D configurations that would all project to the same view.
We refer to this non-uniqueness as \emph{multi-solutionness}.
Typical reconstruction pipelines treat underdetermination as error or instability to be regularized away, collapsing the solution set into a seemingly decisive mesh, depth map, or radiance field.

Multi-solutionness is not merely a technical nuisance; it is where a painting's spatial imagination resides, and where viewers' inferences can legitimately diverge.
We distinguish two sources.
First, \textbf{unobserved content} (backsides, occluded volumes, and out-of-frame continuations) must be invented beyond the depicted viewpoint, and multiple inventions can remain pictorially plausible.
Second, \textbf{observed cues} do not uniquely constrain 3D geometry: perspective and lighting admit families of solutions (e.g., bas-relief ambiguity \citep{belhumeur1999basrelief}).

To make both levels observable, we propose a three-stage workflow that back-translates a painting into a \emph{set} of physical 3D artifacts instead of a single converged result.
We first sample multiple camera-orbit (turntable) multi-view video sequences with a video generative model; each sequence instantiates a different hypothesis about unobserved content.
For each sequence, we reconstruct a 3D Gaussian Splatting representation \citep{kerbl2023gaussian} whose point density, ghosting, and boundary diffusion retain disagreements across synthesized frames, exposing where observed cues still leave room for interpretation.
Finally, we leverage DreamPrinting \citep{wang2025dreamprinting} to fabricate these reconstructions as physical objects, allowing multi-solutionness to be inspected and compared through embodied viewing.
Our contributions are as follows:
\begin{itemize}
  \item \textbf{A two-level framing of multi-solutionness} in figurative paintings that separates non-uniqueness due to \emph{unobserved content} from non-uniqueness that persists under \emph{observed cues}.
  \item \textbf{A painting-to-3D back-translation pipeline} that treats non-uniqueness as an explicit output, producing multiple hypothesis-conditioned Gaussian-splat reconstructions rather than a single consolidated scene.

  \item \textbf{A materialization step} demonstrating that hypothesis-conditioned 3DGS reconstructions from a single painting can be materialized and compared in physical form using DreamPrinting as a downstream fabrication backend.
  
\end{itemize}

\begin{figure}[t]
  \centering
  \includegraphics[width=\linewidth]{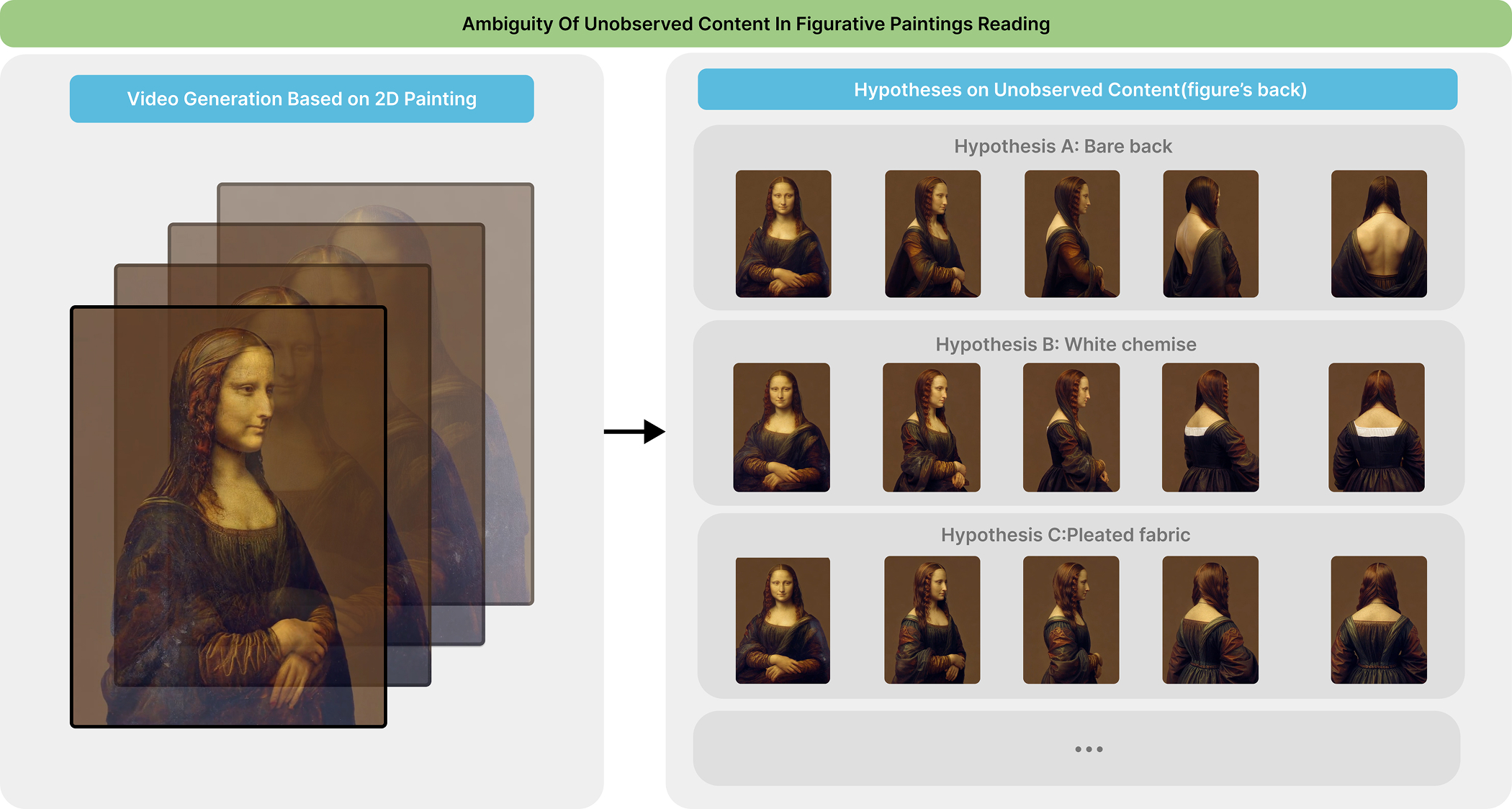}
\caption{\textbf{Unobserved-content multi-solutionness.} From a single figurative painting, multiple image-consistent completions of unseen structure (e.g., the figure’s back) are possible. Sampling an image-conditioned video generator yields different camera-orbit sequences that preserve the input view but diverge in unobserved content, producing distinct spatial and semantic hypotheses for subsequent reconstruction and fabrication.}
  \label{fig:unobserved}
\end{figure}

\section{Related Works}
Art historical and perceptual research has long emphasized that figurative paintings require the viewer to reconstruct a spatial layout that is inherently underdetermined\citep{willats1997art, hertzmann2010non}. Vision is intrinsically ambiguous, and art deliberately exploits this ambiguity through conventions that shape how depth, illumination, and spatial relations are read\citep{mamassian2008ambiguities, cavanagh2005artist}. Consequently, rather than yielding a single ``correct'' geometry, pictorial depth can remain compatible with multiple valid interpretations, functioning as a ``dialectical device'' between the flat surface and the depiction of deep space \citep{biberman2017insideout, florensky2002reverse}.

In contrast, standard computer vision pipelines treat this ill-posed nature as a problem to be resolved by optimizing for a single, consolidated scene. Techniques such as Structure-from-Motion and Multi-View Stereo \citep{schoenberger2016sfm,furukawa2010mvs}, as well as Neural Radiance Fields and 3D Gaussian Splatting \citep{mildenhall2020nerf,kerbl2023gaussian}, typically aim for convergence to one coherent instance. While recent work explicitly incorporates uncertainty \citep{hoffman2023probnerf,pan2022activenerf}, residual ambiguity is still commonly treated as an internal error term rather than an interpretable output. Generative systems offer a different route by using learned priors to propose plausible 3D content; however, both optimization-based methods \citep{poole2022dreamfusion,lin2023magic3d,wang2023prolificdreamer} and feed-forward 3D generators \citep{zhang2024clay,yang2024hunyuan3d,tochilkin2024triposr} often re-synthesize appearance toward dataset-canonical renderings, which can obscure the specific material trace of a painting (e.g., brushwork and local texture). Large-scale world models \citep{sun2025worldplay,worldlabs2025marble,openai2024worldsimulators} prioritize scene-level environment generation and long-horizon dynamics , and are not designed for painting-to-3D back-translation where preserving painterly surface qualities is central. Conversely, viewpoint-conditioned novel-view synthesis and modern video generative models \citep{liu2023zero123,blattmann2023svd,qwenImageEdit2511MultiAnglesLora,wan2025} operate closer to pixel space, which can better retain the source painting's texture under viewpoint changes, while often producing view-dependent inconsistencies rather than a strictly unified geometry .

Translating such ambiguous representations into physical matter requires overcoming standard fabrication constraints. Conventional 3D printing generally demands watertight meshes, a requirement that collapses the soft, indeterminate regions one wishes to examine. 

DreamPrinting \citep{wang2025dreamprinting} addresses this by transforming radiance-based volumetric assets into explicit Volumetric Printing Primitives (VPPs), enabling high-fidelity color and volumetric appearance without forcing surface closure. While DreamPrinting provides a powerful technical fabrication backend natively designed for radiance fields, our work introduces a fundamentally different conceptual and data-acquisition pipeline. We treat interpretive multi-solutionness as our core artistic target: we propose our method to extract 3DGS hypotheses from 2D paintings via video generative models, and we adapt the VPP concept to materialize these point-based artifacts.

\begin{figure}[t]
  \centering
  \includegraphics[width=\linewidth]{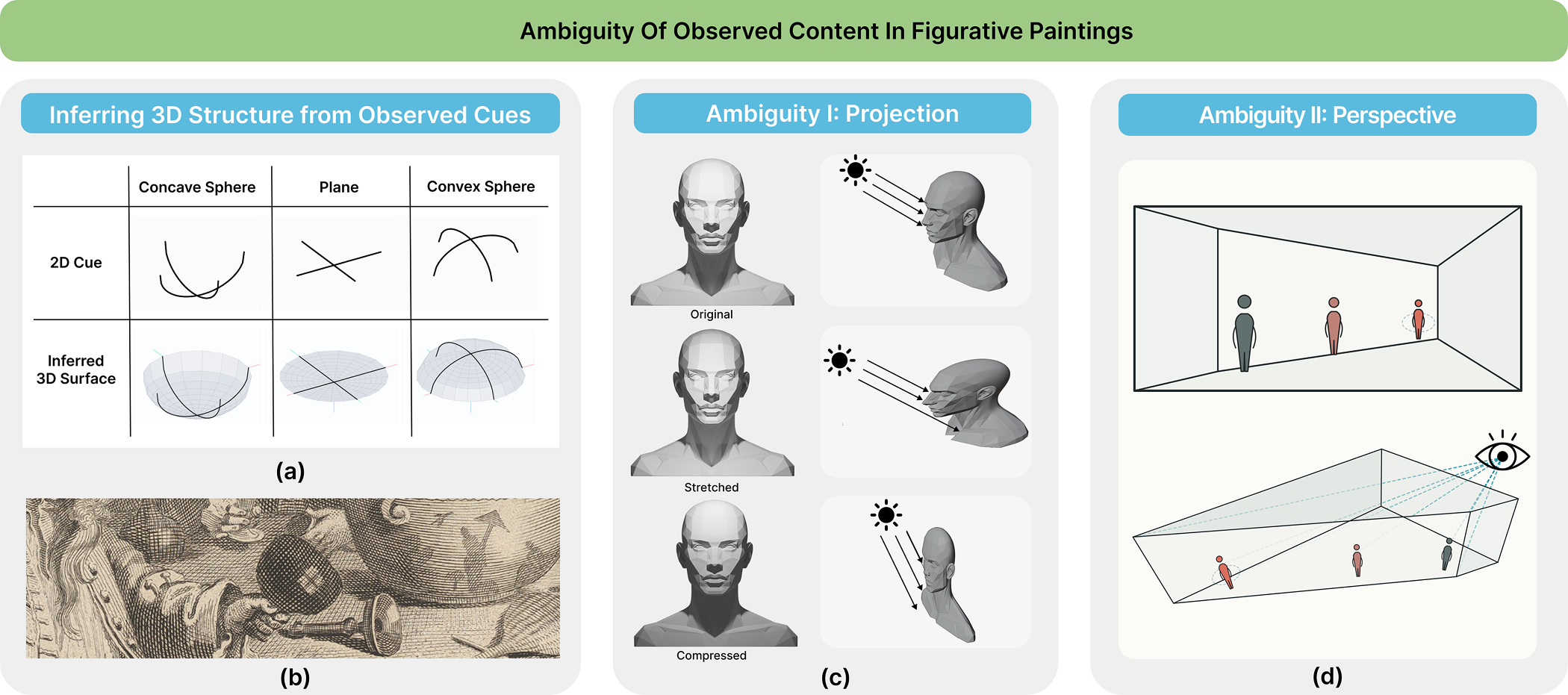}
\caption{\textbf{Observed-cue multi-solutionness.} Even when only visible evidence is considered, pictorial cues underdetermine 3D structure. (a) Demonstrates how 2D curves in an image can correspond to 3D space \citep{mamassian2008ambiguities}. (b) A painting example: the reflection on the glass helps infer its 3D geometry. Detail from Hogarth. \textcopyright~The Metropolitan Museum of Art~\cite{hogarth_midnight_modern_conversation_met}. (c) The bas-relief / shape-from-shading ambiguity, where distinct depth configurations can yield similar image evidence \citep{belhumeur1999basrelief}. (d) The Ames room effect, demonstrating how perspective cues can support incompatible spatial interpretations \citep{Wang2022}.}

  \label{fig:observed}
\end{figure}

\section{Motivation and Concepts}
Most computational approaches to 3D reconstruction are oriented toward producing a single, coherent spatial result, implicitly treating ambiguity as a defect to be eliminated. In doing so, they collapse multiple compatible interpretations, long acknowledged in art-historical discourse, into a singular spatial claim, effectively replacing interpretive openness with algorithmic resolution.

Our workflow is designed to make \emph{multi-solutionness} observable in two complementary ways: across alternative hypotheses about \emph{unobserved content} (what the painting does not specify from its single viewpoint), and within each hypothesis through how ambiguity in \emph{observed cues} (projection, perspective, shading, occlusion) persists when a sequence is treated as multi-view evidence (Figure~\ref{fig:unobserved} and~\ref{fig:observed}).

\subsection{Unfolding multi-solutionness through generative multi-view synthesis}
We first employ an image-conditioned video generative model to synthesize multiple camera-orbit (turntable-style) multi-view image sequences from a s ingle figurative painting. Each sequence represents one plausible completion of the \emph{unobserved content} implied by the original image, such as backsides, occluded structure, or out-of-frame continuation. While preserving the pictorial constraints of the source painting, generating multiple sequences makes the space of compatible completions explicit: each set of views constitutes a legitimate yet non-unique interpretation of what the painting leaves unseen.

Within an individual sequence, the newly generated views are inferred from the visible evidence in the painting and from the frames generated so far. Temporal consistency in video generation organizes the frames into a coherent orbit around a single hypothesized scene. Importantly, this coherence is driven by cue propagation: each generated frame becomes the next step's visual evidence, so its pixels and relations function as the \emph{observed cues} for subsequent frames. In this way, the sequence provides multi-view evidence for reconstruction, while also carrying forward the residual ambiguity of observed cues that cannot be uniquely resolved by a single image.

\subsection{Preserving multi-solutionness in 3D Gaussian Splatting reconstruction}
Subsequently, we apply 3D Gaussian Splatting (3DGS)~\citep{kerbl2023gaussian} to each synthesized multi-view sequence independently. Each reconstructed 3DGS corresponds to one hypothesis about \emph{unobserved content} unfolded from the same painting, yielding a set of hypothesis-conditioned 3D representations rather than a single consolidated scene.

During reconstruction, 3DGS aligns multi-view cues by optimizing a point-based Gaussian scene representation, namely a set of 3D Gaussian primitives rendered by splatting. Unlike mesh-based pipelines that tend to force a hard boundary decision, 3DGS can retain overlapping support and soft transitions. As a result, regions where multi-view evidence is consistent concentrate into sharper structures, while unresolved regions appear as diffuse bands, boundary softening, or ghosted duplicates. In our workflow, these structures are not treated as errors to be suppressed; they are the visible signature of observed-cue ambiguity persisting within a fixed unobserved-content hypothesis.

\begin{figure}[t]
  \centering
  \includegraphics[width=0.9\linewidth]{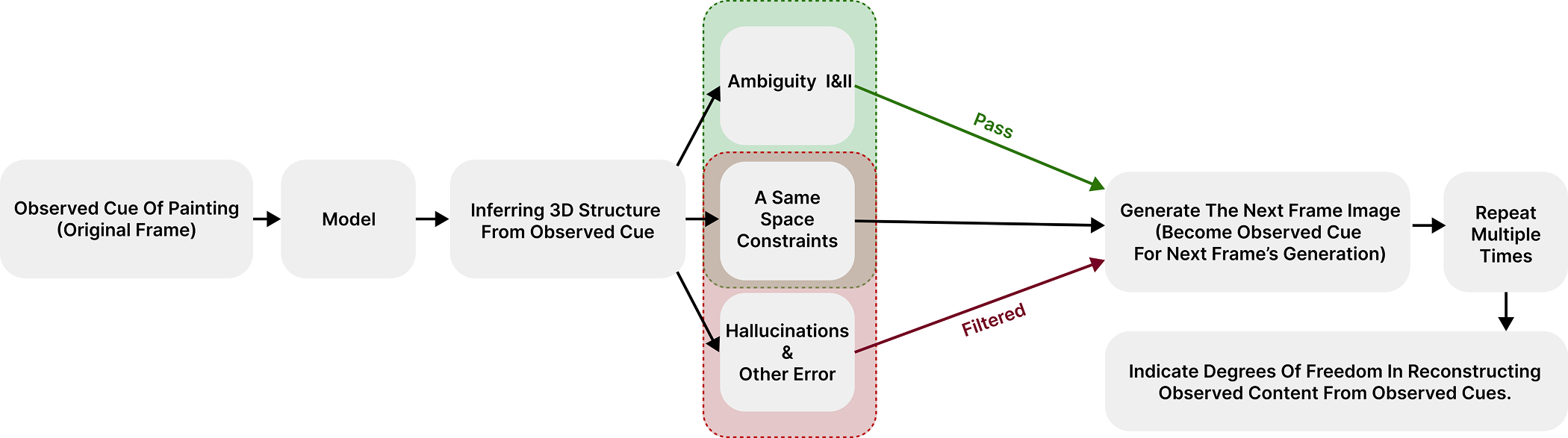}
\caption{\textbf{Observed-cue propagation in image-conditioned video synthesis.} Starting from the original frame, the model infers a provisional 3D structure from observed cues and uses it to generate the next-frame image, which then becomes the observed cue for the subsequent step. Repeating this cue-to-image update under the same-space constraints yields a temporally coherent orbit while revealing degrees of freedom in reconstructing observed content from observed cues; deviations that violate the constraints appear as hallucinations and are filtered or suppressed during generation.}
  \label{fig:observed2}
\end{figure}

\subsection{Materializing multi-solutionness}
Finally, we materialize the reconstructed results as physical artifacts. Digital splat reconstructions are typically read through interactive navigation, where indeterminate structure can be overlooked or reinterpreted as the viewpoint changes. By contrast, physical objects provide persistent spatial presence that supports inspection and comparison.

We directly leverage DreamPrinting \citep{wang2025dreamprinting} as our manufacturing medium. While originally demonstrated on continuous neural radiance fields, we adapt its Volumetric Printing Primitives to convert our generated 3DGS data into physical matter without forcing watertight surface closure.

Materialization here is not intended to resolve non-uniqueness; instead, it stabilizes it. The resulting set of prints externalizes different unobserved-content hypotheses, while each individual print retains the soft boundaries and overlaps that mark where observed cues remain compatible with multiple 3D solutions.

\section{Implementation}

\subsection{Overview and notation}
We take as input a single RGB painting $I\in\mathbb{R}^{H\times W\times 3}$ and produce a set of point-based 3D reconstructions $\{\mathcal{G}_k\}_{k=1}^K$, where each $\mathcal{G}_k$ is a set of anisotropic 3D Gaussians with color and opacity parameters as in 3D Gaussian Splatting \citep{kerbl2023gaussian}.
Each reconstruction is associated with a synthesized multi-view sequence $V_k=\{I^t_k\}_{t=1}^{M}$.

\subsection{Stage 1: Generative multi-view synthesis}
We use an image-to-video generative model to synthesize turntable-style sequences around the painting's central subject.
Concretely, we employ Jimeng Video 3.5 Pro \citep{jimengai} as an off-the-shelf generator.
We condition the model on the input painting and a prompt specifying a slow, approximately $360^\circ$ orbit while preserving the original composition and lighting as much as possible.
To sample different hypotheses about unobserved content, we generate $K$ sequences by varying the random seed and (when supported) prompt variations that keep the same pictorial constraints.

We denote the sampling process as
\begin{equation}
  V_k \sim p_{\theta}\!\left(V \mid I, \tau, z_k\right),
\end{equation}
where $\tau$ is a fixed prompt template and $z_k$ controls stochasticity.
The resulting set $\{V_k\}$ externalizes unobserved-content multi-solutionness as multiple compatible completions.

\subsection{Stage 2: Multi-view Gaussian Splatting reconstruction}
For each sequence $V_k$, we reconstruct a 3D Gaussian scene representation $\mathcal{G}_k$.

\paragraph{Camera estimation.}
We estimate per-frame camera poses using COLMAP \citep{colmap}, treating the generated frames as a multi-view set.
While camera estimation from synthesized imagery is imperfect, in our context pose noise is not purely detrimental: it contributes to the density-based visualization of underconstrained structure.

\paragraph{Foreground masking.}
To focus reconstruction on the depicted subject and avoid background ambiguities dominating the optimization, we compute per-frame alpha mattes with MatAnyone \citep{yang2025matanyone}.
Let $A^t_k\in[0,1]^{H\times W}$ denote the matte for frame $t$.

\paragraph{Optimization.}
Given camera parameters $C_k$ and mattes $A_k$, we optimize $\mathcal{G}_k$ via differentiable splatting, rendering an image $\hat{I}^t_k = \mathcal{R}(\mathcal{G}_k, C^t_k)$ for each view.
We apply a foreground photometric term and a background transparency term:
\begin{align}
  \mathcal{L}_{\text{fg}} &= \sum_{t}\left\lVert A^t_k \odot \left(\hat{I}^t_k - I^t_k\right)\right\rVert_{1},\\
  \mathcal{L}_{\alpha} &= \sum_{t}\left\lVert (1-A^t_k) \odot \alpha\!\left(\mathcal{R}(\mathcal{G}_k, C^t_k)\right)\right\rVert_{1},\\
  \mathcal{L} &= \mathcal{L}_{\text{fg}} + \lambda\,\mathcal{L}_{\alpha},
\end{align}
where $\alpha(\cdot)$ extracts the rendered opacity and $\lambda$ balances terms.
Following \citet{kerbl2023gaussian}, we use standard 3DGS parameter updates (positions, covariances, colors, opacities) to fit the synthesized sequence.
To preserve observed-cue multi-solutionness, we avoid aggressive regularization that would collapse diffuse regions into a sharp surface.

\subsection{Stage 3: Volumetric fabrication via DreamPrinting}

Point-based radiance representations are not directly printable, and conventional meshing can erase the boundary indeterminacy we seek to preserve. We therefore leverage DreamPrinting \citep{wang2025dreamprinting} as our fabrication backend. Because its original implementation targets continuous radiance fields, we extend its radiance-to-print pipeline to process the discrete density and color attributes of our optimized 3DGS representations ($\mathcal{G}_k$). By converting 3DGS into Volumetric Printing Primitives, we output printer-ready pigment labels in \texttt{CMYKWCl} for a Stratasys J850 Prime PolyJet printer. We fabricate the results without forcing watertight surface closure, preserving boundary softness and density variation as material properties of the printed object.

\begin{figure}[t]
  \centering
  \includegraphics[width=\linewidth]{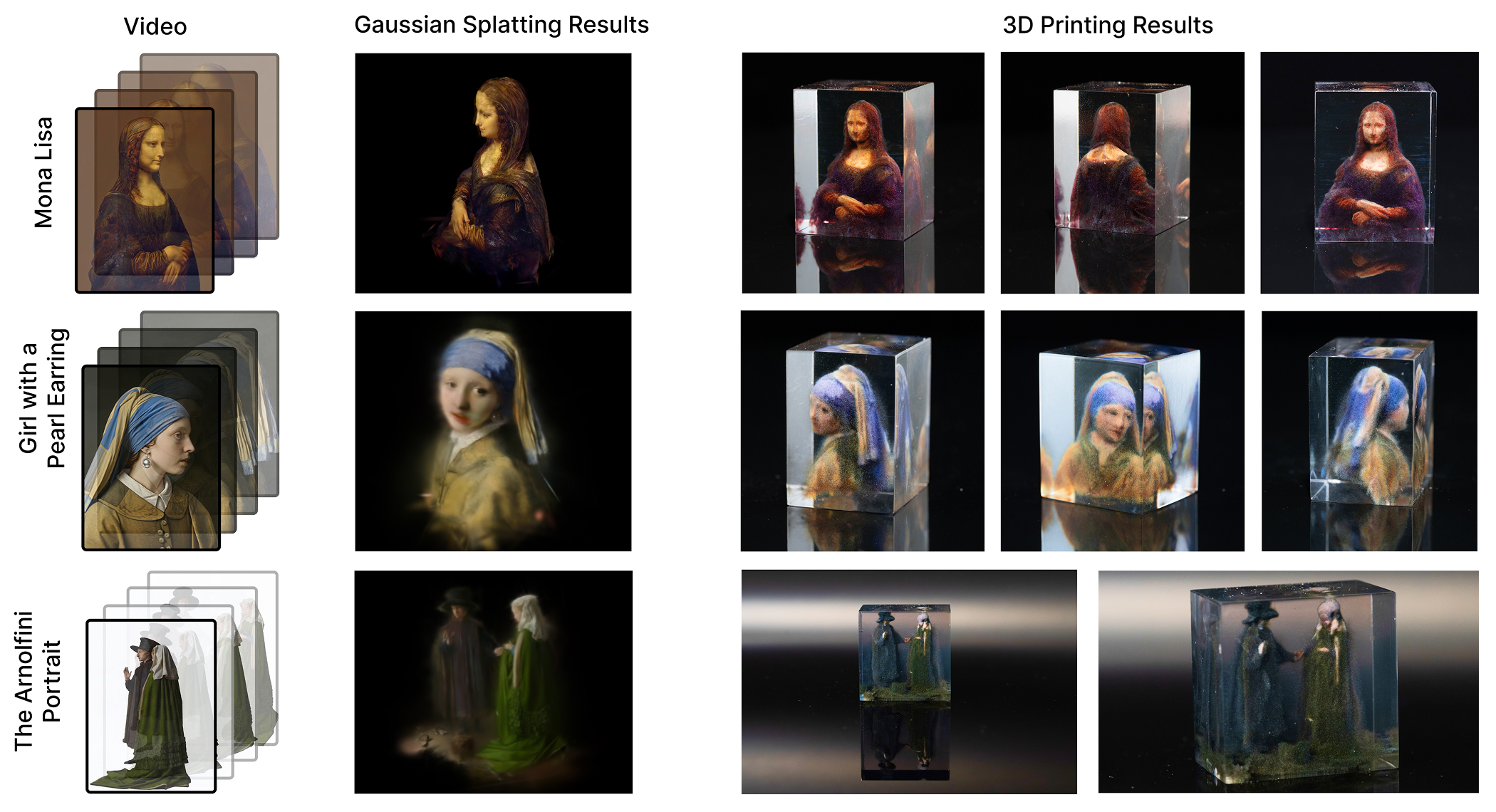}
\caption{\textbf{Physicalizing multi-solutionness from paintings.} For each input painting, we reconstruct a hypothesis-conditioned 3D Gaussian Splatting representation and fabricate it as a volumetric print via DreamPrinting. The printed artifacts preserve the painting’s appearance while making spatial non-uniqueness inspectable as a physical object, including softened boundaries and ghosted structure where observed cues remain underconstrained. Source paintings: \textcopyright{} public-domain images via Wikimedia Commons~\cite{commons_public_domain_paintings_fig5}.}

  \label{fig:results}
\end{figure}
 
\section{Discussion and Future Work}
\subsection{Artistic Significance of materialized uncertainty}
First, unobserved-content multi-solutionness explicitly engages the implied off-screen space. Excavating these hidden perspectives is a critical pathway to understanding and completing the structural logic of pictorial representation \citep{searle1980las}. By generating physical hypotheses for the unpainted, the workflow transforms the canvas from a flat boundary into a speculative spatial narrative.

Second, observed-content multi-solutionness materializes visual ambiguity as varying physical density, where structural diffuseness serves as a visual index of multi-view cue consistency. This creates a tangible "possibility space" for interpretation. Meantime, preserving this geometric blur cultivates a unique expressive power. Rather than collapse a painting into a single form, unresolved visual cues actively invite the viewer’s cognitive participation and psychological projection, evoking a distinct affective resonance \citep{cavanagh2005artist}. Within this realm, the artifact itself emerges as a novel lens for viewing art history. It allows viewers to physically experience pictorial spaces as tangible artistic archives.

The essential value of this artistic practice is then established: these physical artifacts do not merely reproduce paintings; they materialize the very act of viewing and questioning. By allowing multiple compatible readings to coexist simultaneously, they transform the abstract, often invisible process of art-historical inquiry into a tangible, navigable spatial experience.

\subsection{Limitations}
When a painting lacks a clear subject or anchor, the model struggles to synthesize coherent rotational sequences, leading to fragmented reconstructions. 

Furthermore, Gaussian Splatting preserves these synthesized inconsistencies as diffuse structures, unable to distinguish generative artifacts from meaningful spatial ambiguity. 

Consequently, we treat the final outputs as heuristic artifacts rather than ground-truth spatial recoveries. In this context, assessing the visual similarity between the physical object and the source image when observed from the painting's original reference viewpoint, serves as the primary method for evaluating the quality of a printed artifact.

A second limitation is representational integration. Currently, unobserved-content multi-solutionness requires multiple distinct printed objects, whereas observed-content multi-solutionness is expressed as density variation within each object. 

Encoding both dimensions of uncertainty into a single, unified 3D-printed sculpture without losing analytical comparability remains an open direction.

\subsection{Future work}
Future work could study how different pictorial styles distribute constraints and degrees of freedom, e.g., whether certain compositional strategies lead to characteristic density patterns or ghosting modes in the reconstructed 3DGS representations.
On the system side, integrating explicit multi-view consistency checks could enable controlled adjustment of how much drift is retained, making multi-solutionness legible at different granularities.
Finally, extending the workflow to non-photographic media beyond painting (illustration, printmaking, stylized digital art) could help map the boundaries of multi-solutionness across representational systems.

\section{Conclusion}
We presented a workflow that materializes multi-solutionness in figurative paintings (Figure~\ref{fig:results}).
By sampling multiple generative view sequences, reconstructing each with Gaussian Splatting, and fabricating the resulting Gaussian-splat reconstructions via DreamPrinting, the method treats non-uniqueness as an explicit output rather than an error to be suppressed.
The resulting digital and physical artifacts provide a practical reference for examining how 2D depiction, spatial inference, and generative systems interact, opening directions for art-oriented investigation of uncertainty and distributional structure in spatial representation.

\begin{acks}
The authors thank Prof. Jie Wang and Prof. Hua Yang for their valuable advice and guidance on this work. This work was also supported in part by the
\grantsponsor{NSFC}{National Natural Science Foundation of China}{}
under Grant~\grantnum{NSFC}{W2431046}, \grantsponsor{NKRDPC}{National Key R\&D Program of China}{} \grantnum{NKRDPC}{2025YFA1309603}, the
\grantsponsor{CGLSTF}{Central Guided Local Science and Technology Foundation of China}{}
under Grant~\grantnum{CGLSTF}{YDZX20253100001001}, and by the
MoE Key Lab of Intelligent Perception and Human-Machine Collaboration (ShanghaiTech University),
the Shanghai Frontiers Science Center of Human-centered Artificial Intelligence,
and the HPC Platform of ShanghaiTech University.
\end{acks}

\bibliographystyle{ACM-Reference-Format}
\bibliography{references}

\end{document}